\documentclass[aps,prl,preprint,groupedaddress]{revtex4-2}

\usepackage{amsmath}
\usepackage{graphicx}
\begin{document}


\title{Extended Mean-Field Theory for the 2D Hubbard Model in Degenerate Dilute Electron Gases: Fluctuations, Superconducting Dome, and Interaction Mechanisms in Strontium Titanate}


\author{Xing Yang}
\email[]{xyang@guet.edu.cn}
\affiliation{School of Materials Science and Engineering, Guilin University of Electronic Technology, Guilin, Guangxi Zhuang Autonomous Region, China, 541000}

\author{Xinyu Zhang}
\affiliation{School of Materials Science and Engineering, Guilin University of Electronic Technology, Guilin, Guangxi Zhuang Autonomous Region, China, 541000}

\author{Xuchang Zhang}
\affiliation{School of Materials Science and Engineering, Guilin University of Electronic Technology, Guilin, Guangxi Zhuang Autonomous Region, China, 541000}


\date{\today}

\begin{abstract}
Strontium titanate ($\mathrm{SrTiO_3, STO}$) dome-shaped superconducting transition temperature as a function of chemical potential, consistent with STO experiments, and shows that tunable s-wave and d-wave symmetries are modulated by doping. Superconducting fluctuations validate the mean-field approximation at low temperatures but destroy pairing at higher temperatures. The charge-density-wave order competes with superconductivity, enhances the effective electron mass inversely with the chemical potential, and increases with the interaction strength $U$ and the temperature $T$. SDW order is rare and fragile, while an additional magnetic term induces subtle band splitting. These findings suggest e-e contributions to STO's transport anomalies and provide criteria to distinguish e-e from e-ph origins, offering insights for engineering higher $T_c$ in dilute systems.
\end{abstract}

\maketitle

\section{Introduction}

Strontium titanate ($\mathrm{SrTiO_3}$, STO) stands as a paradigmatic dilute superconductor, first identified in 1964 with a maximum superconducting transition temperature ($T_c$) of approximately 0.3 K in oxygen-deficient samples, characterized by extremely low carrier densities ($n \sim 10^{17}-10^{20} \mathrm{cm^{-3}}$) and a Fermi energy ($\epsilon_F \sim 1-10 meV$) comparable to or below phonon energies\cite{PhysRevLett.12.474}. Over the past decade, significant advancements have illuminated its unconventional superconductivity, driven by innovations in doping strategies\cite{doi:10.1126/sciadv.abl5668}, strain engineering\cite{PhysRevB.111.054522}, and nanoscale probing techniques\cite{christensen_2024_2024}. Notable progress includes the observation of enhanced $T_c$ near quantum critical points (QCPs) in electron-doped systems, such as La-substituted $\mathrm{Sr_{1-x}La_xTiO_3}$, where superconductivity emerges proximate to non-magnetic instabilities and ferroelectric quantum criticality\cite{zhang_enhanced_2025}. In 2024, studies revealed metallic and superconducting states induced by thermal or electrical deoxidation of surface dislocation networks in STO, yielding $T_c$ enhancements through filamentary bundles and polaronic coupling\cite{nano14231944}. Furthermore, later research documented polarization-density waves in STO, manifesting as soft polar-acoustic collective modes at nanometer scales and detected via femtosecond X-ray and terahertz spectroscopy, underscoring the role of mesoscopic modulation near QCPs in boosting superconducting pairing\cite{orenstein_observation_2025}. A recent study highlighted ferroelastic domain walls as preferential sites for quasi-1D superconductivity, with $T_c$ variations mapped across twin structures in $\delta$-doped STO, achieving localized enhancements\cite{ojha2026electronsmeetferroelasticdomain}. The interfacial systems, such as FeSe/STO heterostructures, demonstrated strong electron-phonon (e-ph) coupling, elevating $T_c$ to ~40-110 K\cite{yang_phonon_2024}. Coexisting superconductivity and ferroelectricity were reported in 2024 at $\mathrm{LaAlO_3/KTaO_3}$ interface with $T_c$ boosted by 0.2-0.6 K via polarization-induced confinement\cite{tomioka_enhanced_2019}.

Central to these advances is the ongoing debate over superconductivity mechanisms, pitting electron-phonon (e-ph)\cite{phn} against electron-electron (e-e) interactions \cite{saha_strong_2025, plm}. The e-ph mechanism is supported by interactions with longitudinal optical (LO) phonons (frequencies $\sim 10-100 meV$), particularly in the dilute regime where $\epsilon_F$ is below phonon energies\cite{doi:10.1073/pnas.1604145113}. A far-infrared study of metallic STO revealed spin-orbit-assisted e-ph coupling, manifesting as dynamic Rashba spin-orbit interactions that hybridize plasmons, phonons, and electron spin-flip modes under magnetic fields, with observed soft TO phonon modes\cite{PhysRevB.108.075162}. Generalized Rashba e-ph coupling models have been developed for STO's dome-shaped $T_c$ vs. doping, attributing this to multi-band effects and proximity to ferroelectric phases \cite{PhysRevResearch.5.023177}. Strong-coupling theories in integrated ferroelectric quantum criticality, in which soft TO-$\Gamma$ modes enhance e-ph pairing, predicting that $T_c$ increases near QCPs\cite{saha_strong_2025}. Quadratic e-ph coupling promotes quantum bipolaron formation, potentially elevating $T_c$ in low-dimensional systems like STO interfaces\cite{PhysRevLett.132.226001}. Conversely, e-e mechanisms invoke plasmon-mediated pairing or Hubbard-like correlations, particularly in strongly coupled regimes, where plasmons (energies could facilitate attraction\cite{in_t_Veld_2023}, though recent analyses deemed them insufficient alone, often requiring disorder or mobility edges for net attraction\cite{moroder_phonon_2024}. Isotope effects ($^{18}O$ substitution shifts the $T_c$ dome) and phonon-state tomography revealed e-ph-driven correlation dynamics\cite{stucky_isotope_2016}. However, it is discovered that e-e correlations enhance effective mass \cite{najev_electronic_2025}, which emphasizes the function of the e-e interaction in the formation of superconducting states, and theoretically, the on-site e-e interaction can exist in materials (e.g., STO) even with a high dielectric constant.

Electron transport in dilute STO has undergone parallel refinements, emphasizing density-dependent anomalies and fluctuation effects \cite{PhysRevLett.133.136003}. Metallic STO exhibits enhanced spin susceptibility at low densities, signaling correlations beyond Pauli paramagnetism and coupling to transport via QCP proximity\cite{najev_electronic_2025}. Domain walls induce anisotropic resistivity and glassy relaxations, driven by electron-dipole interactions and strain fields\cite{maity_glassy_2025}. In 2D electron gases (e.g., $\mathrm{LaAlO_3/STO_3}$), Rashba spin-orbit coupling yields spin textures and anomalous Hall conductivity \cite{Zhang_2025}, with photo-induced carriers shifting $T_c$ via tuned SOC\cite{soumyanarayanan_emergent_2016}. These insights not only resolve longstanding mechanistic debates but also pave the way for engineering higher $T_c$ in dilute systems \cite{saha_strong_2025}. However, the role of the on-site e-e interactions in electron transport and superconducting properties remains limited to simulation methodologies \cite{rpa, eta,dft,npe}. 

Previous research on the simulation algorithms of the 2D Hubbard model has advanced significantly, driven by quantum simulation, machine learning, and hybrid quantum-classical algorithms. It has achieved unprecedented accuracy in phase diagrams, ground states, and dynamical properties \cite{article}\cite{PhysRevX.6.031045}\cite{Roth2025}. Although the 2D Hubbard model does not detect superconducting ground states, the inclusion of a next-nearest-neighbor hopping term reveals superconductivity using modernized quantum Monte Carlo methods\cite{doi:10.1126/science.adh7691}. Neural Quantum States (NQS) methods, leveraging transformer architectures, achieved state-of-the-art results in 2025 for doped 2D Hubbard models \cite{gu2025solvinghubbardmodelneural}. Fermionic projected entangled pair states (fPEPS) simulations extended to finite-size 2D Hubbard lattices (up to 20×20), yielding Mott insulator-metal transitions\cite{Liu2025}. Hybrid quantum-classical approaches, such as variational quantum eigensolvers (VQE), benchmarked on classical hardware, simulated Hubbard ground states on noisy intermediate-scale quantum (NISQ) hardware equivalents \cite{doi:10.1126/sciadv.adu9991}. Dynamical mean-field theory (DMFT) integrated with quantum convolutional neural networks classified metal-Mott insulator transitions at thermodynamic limits\cite{quantum7020018}. Extensions to SU(N) Hubbard models in ultracold atoms (2024-2025) revealed exotic correlated phases, such as flavor-ordered insulators for $N>2$, with entropy benchmarks probing quantum advantage\cite{Ibarra_Garc_a_Padilla_2024}. Non-standard Bose-Hubbard variants emulated with superconducting qubits explored entanglement propagation in 4×4 arrays \cite{Karamlou2024Probing}. 

Current research on the 2D Hubbard model has two main weaknesses: for numerical methods, the number of lattice points typically does not meet the requirements to accommodate an electron pair; meanwhile, perturbative methods sometimes diverge at particular k-points. The newly proposed extended mean-field theory (eMFT) approximates the four-operator Hubbard term across three regimes, covers all possible two-operator combinations, and examines the validity of mean-field approximations using numerical methods, thereby demonstrating non-perturbative and universal features from weak to strong e-e interactions. Although superconductivity has been detected in the 2D Hubbard model using eMFT, in contrast to quantum Monte Carlo algorithms that disproved the superconductivity, our results did not challenge the previous conclusions. On the contrary, our previous calculations, which incorporate all possible order parameters and solve the Heisenberg equations self-consistently, also fail to find superconductivity. The previous methods are similar to quantum Monte Carlo algorithms; both seek appropriate quasiparticles to diagonalize the 2D Hubbard model Hamiltonian, but one uses mean-field approximations, and the other uses the Hubbard-Stratonovich transformation. The absence of superconductivity, as deduced by both algorithms, may be due to the complex structures of multiple order parameters across different parameter spaces and k-space, which make finding finite fixed points infeasible, although reaching the final conclusions requires further exploration.

\section{Theoretical Framework: 2D Hubbard Model and Extended Mean-Field Theory}
Let us start with a 2D degenerate gas with a Hubbard term \cite{mahan},

\begin{equation}
\hat{H}=\hat{H}_K+\hat{H}_U
\end{equation}

\noindent where $\hat{H}, \hat{H}_K, \hat{H}_U$ are the total, kinetic part of, and Hubbard term of the Hamiltonian. The kinetic part is
\begin{equation}
\hat{H}_K=\sum_{\mathbf{k}, \sigma}E_{\mathbf{k}} \hat{c}^\dagger_{\mathbf{k}\sigma} \hat{c}_{\mathbf{k}\sigma}
\end{equation}

\noindent where $E_{\mathbf{k}} = \frac{\hbar^2 |\mathbf{k}|^2}{2m}-\mu$ is the kinetic energy of the electron with wave vector $\mathbf{k}$, mass $m$, and chemical potential $\mu$. $ \hat{c}^\dagger_{\mathbf{k}\sigma} (\hat{c}_{\mathbf{k}\sigma})$ are the creation (annihilation) operators of the electron with wave vector $\mathbf{k}$ and spin $\sigma$. The Hubbard term is

\begin{equation}
\hat{H}_U=\sum_{i} U \hat{n}_{i \uparrow} \hat{n}_{i \downarrow} 
\end{equation}

\noindent where $\hat{n}_{i \sigma}=\hat{c}^\dagger_{\mathbf{i} \sigma}\hat{c}_{\mathbf{i} \sigma}$ is the electron number operator on site $\mathbf{i}$ with spin $\sigma$, and $ \hat{c}^\dagger_{\mathbf{i}\sigma} (\hat{c}_{\mathbf{i}\sigma})$ are the creation (annihilation) operators of the electron on site $\mathbf{i}$ with spin $\sigma$. With the Fourier transformation, $\hat{c}^{(\dagger)}_{\mathbf{k}\sigma}=\frac{1}{\sqrt{N}}\sum_{\mathbf{i}} e^{-i \mathbf{k}\cdot  \mathbf{i}}\hat{c}^{(\dagger)}_{\mathbf{i}\sigma}$, the Hubbard term exhibits the conservation of electron momenta during the scattering process, and one of the algebraic forms of the Hubbard term is

\begin{equation}
\hat{H}_U=\sum_{\mathbf{k},\mathbf{k'},\mathbf{q}} \frac{U}{N}  \hat{c}^\dagger_{\mathbf{k}+\mathbf{q}/2,\uparrow} \hat{c}^\dagger_{-\mathbf{k}+\mathbf{q}/2,\downarrow} \hat{c}_{-\mathbf{k'}+\mathbf{q}/2,\downarrow} \hat{c}_{\mathbf{k'}+\mathbf{q}/2,\uparrow}
\end{equation}

\noindent where $N$ is the total number of quantum states. Solving the Hamiltonian $\hat{H}$ with the perturbation theory frequently results in divergent solutions, especially in the intermediate regime. The strong correlations in the expansion series mainly cause the divergence or enlarge the convergence radius. The newly proposed eMFT offers an effective strategy that treats strong correlations within mean-field theory and diagonalizes them together with the kinetic terms. Therefore, the remaining interaction terms with weak correlations are perturbative\cite{phd}. In the specific calculations, solving two problems is crucial for validating the eMFT: one concerns the conditions under which the mean-field approximation applies, and the other concerns how to calculate the system with multiple strong-correlation terms. The first one can be settled by computing the fluctuations $F$ of a specific strong correlation $C$; when $\bar{F} <  \bar{C}$, the mean-field approximation is valid ($\bar{\hat{Q}} $ is the thermal average of a quantum operator $\hat{Q}$). The other problem can be addressed by the following procedure: the Hubbard term, approximated with the mean-field approach, generates three possibilities:

\begin{equation}
\hat{H}_\Delta=\sum_{\mathbf{k},\mathbf{k'},\mathbf{q}} \frac{U}{2N} (\langle \hat{c}^\dagger_{\mathbf{k}+\mathbf{q}/2,\uparrow} \hat{c}^\dagger_{-\mathbf{k}+\mathbf{q}/2,\downarrow} \rangle\hat{c}_{-\mathbf{k'}+\mathbf{q}/2,\downarrow} \hat{c}_{\mathbf{k'}+\mathbf{q}/2,\uparrow}+\langle \hat{c}^\dagger_{\mathbf{k'}+\mathbf{q}/2,\uparrow} \hat{c}^\dagger_{-\mathbf{k'}+\mathbf{q}/2,\downarrow} \langle \hat{c}_{-\mathbf{k}+\mathbf{q}/2,\downarrow} \hat{c}_{\mathbf{k}+\mathbf{q}/2,\uparrow}\rangle)
\label{eq:del}
\end{equation}

\begin{equation}
\hat{H}_L=\sum_{\mathbf{k},\mathbf{k'},\mathbf{q}} \frac{U}{2N} (\langle \hat{c}^\dagger_{\mathbf{k}+\mathbf{q}/2,\uparrow}  \hat{c}_{\mathbf{k'}+\mathbf{q}/2,\uparrow} \rangle\hat{c}^\dagger_{-\mathbf{k}+\mathbf{q}/2,\downarrow} \hat{c}_{-\mathbf{k'}+\mathbf{q}/2,\downarrow}+\hat{c}^\dagger_{\mathbf{k}+\mathbf{q}/2,\uparrow}  \hat{c}_{\mathbf{k'}+\mathbf{q}/2,\uparrow} \langle\hat{c}^\dagger_{-\mathbf{k}+\mathbf{q}/2,\downarrow} \hat{c}_{-\mathbf{k'}+\mathbf{q}/2,\downarrow}\rangle)
\end{equation}

\begin{equation}
\begin{split}
\hat{H}_M
&= \sum_{\mathbf{k},\mathbf{k'},\mathbf{q}} \frac{U}{2N} \Bigg(
    \langle \hat{c}^\dagger_{\mathbf{k}+\mathbf{q}/2,\uparrow} \hat{c}_{-\mathbf{k'}+\mathbf{q}/2,\downarrow} \rangle
    \hat{c}^\dagger_{-\mathbf{k}+\mathbf{q}/2,\downarrow} \hat{c}_{\mathbf{k'}+\mathbf{q}/2,\uparrow} + \hat{c}^\dagger_{\mathbf{k}+\mathbf{q}/2,\uparrow} \hat{c}_{-\mathbf{k'}+\mathbf{q}/2,\downarrow}
    \langle \hat{c}^\dagger_{-\mathbf{k}+\mathbf{q}/2,\downarrow} \hat{c}_{\mathbf{k'}+\mathbf{q}/2,\uparrow} \rangle
\Bigg) \\
&\quad + U \sum_{\mathbf{k}} \hat{c}^\dagger_{\mathbf{k}\uparrow} \hat{c}_{\mathbf{k}\uparrow}.
\end{split}
\label{eq:sdw}
\end{equation}

Electrons with wave vector $\mathbf{k}$ can be described by one of the Hamiltonians above, determined by calculating fluctuations in the order parameters $\Delta^{(\dagger)}_\mathbf{q} \equiv \langle\hat{c}^{(\dagger)}_{-\mathbf{k}+\mathbf{q}/2,\downarrow} \hat{c}^{(\dagger)}_{\mathbf{k}+\mathbf{q}/2,\uparrow}\rangle$, $L _{\mathbf{k}+\mathbf{q}/2,\mathbf{k'}+\mathbf{q}/2,\sigma} \equiv \langle \hat{c}^\dagger_{\mathbf{k}+\mathbf{q}/2,\sigma}  \hat{c}_{\mathbf{k'}+\mathbf{q}/2,\sigma} \rangle$, $M_{\mathbf{k}+\mathbf{q}/2,-\mathbf{k'}+\mathbf{q}/2, \uparrow} \equiv \langle \hat{c}^\dagger_{\mathbf{k}+\mathbf{q}/2,\uparrow} \hat{c}_{-\mathbf{k'}+\mathbf{q}/2,\downarrow} \rangle$, $M_{-\mathbf{k}+\mathbf{q}/2,\mathbf{k'}+\mathbf{q}/2, \downarrow}  \equiv  \langle \hat{c}^\dagger_{-\mathbf{k}+\mathbf{q}/2,\downarrow} \hat{c}_{\mathbf{k'}+\mathbf{q}/2,\uparrow} \rangle$. $\Delta^{(\dagger)}_\mathbf{q}$ is the (conjugate of) superconducting order parameter\cite{mahan}. $L _{\mathbf{k}+\mathbf{q}/2,\mathbf{k'}+\mathbf{q}/2,\sigma}$ denotes the charge wave density order parameter\cite{cdw}. $M_{\mathbf{k}+\mathbf{q}/2,\mathbf{k'}+\mathbf{q}/2, \uparrow}$ represents the Fourier-transverse magnetization or spin-density-wave order parameter at wave vector $\mathbf{q}$ \cite{sdw}, and the extra term $U \sum_{\mathbf{k}} \hat{c}^\dagger_{\mathbf{k}\uparrow} \hat{c}_{\mathbf{k}\uparrow}$ can induce a finite magnetization in $z$-axis. Using the self-consistent methods in Appendix A, the order parameters are obtained, and they all approach zero. This indicates that states with multiple order parameters have not been discovered. However, the self-consistent methods (see Appendix B) yield finite solutions, which satisfy the validity condition of the mean-field approximations.

\section{Superconducting Properties: Dome-Shaped Transition Temperature, Gap Fluctuations}

\begin{figure}
\includegraphics[width=0.8\textwidth]{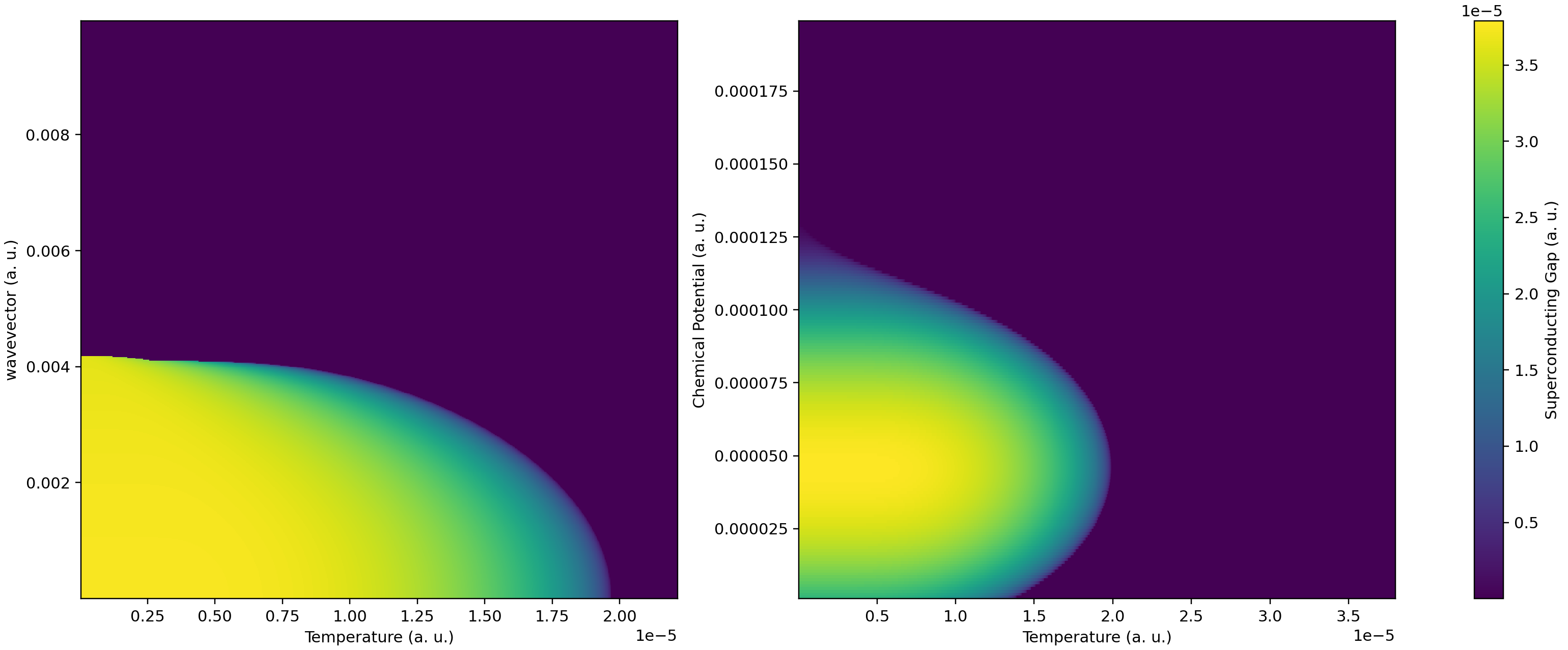}%
\caption{The relation between the superconducting gap and the wave vector (left panel) and the chemical potential (right panel) is plotted at $U= 0.005 ~\mathrm{a. u.}$}
\label{fig:sc}
\end{figure}

The relations of the superconducting gap $\Delta_{\mathbf{q}}$ vs. the chemical potential $\mu$ and wave vector $\mathbf{q}$ are plotted in Fig. (\ref{fig:sc}). The right image shows the superconducting dome modulated by an increase in the chemical potential, consistent with experimental data on superconducting transition temperatures as a function of carrier density, and the magnitude of the superconducting gap in STO. The left image describes the relation between the superconducting gap and the magnitude of wave vector $\mathbf{q}$, which illustrates that the superconducting gap monotonically decreases with the rise of the magnitude of wave vector$\mathbf{q}$. Ref. \cite{phd} demonstrates that the critical wave vector $\mathbf{q}_c$ where the superconducting gap diminishes can result in a peak-dip-hump feature in cuprates.

Additionally, the validity of the mean-field approximations needs to be examined by evaluating the magnitude of superconducting gap fluctuations, which is

\begin{equation}
F(\hat{\Delta}_{\mathbf{q}})=\sum_{\mathbf{k},\mathbf{k'}}  \langle   \hat{c}^\dagger_{\mathbf{k}+\mathbf{q}/2,\uparrow} \hat{c}^\dagger_{-\mathbf{k}+\mathbf{q}/2,\downarrow} \hat{c}_{-\mathbf{k'}+\mathbf{q}/2,\downarrow} \hat{c}_{\mathbf{k'}+\mathbf{q}/2,\uparrow}   \rangle - C(\Delta_{\mathbf{q}})
\end{equation}

where $C(\Delta_{\mathbf{q}})=  \langle  |\Delta_{\mathbf{q}}|^2   \rangle$, $F(\hat{\Delta}_{\mathbf{q}})$ are the first and second moment of the superconducting gap $\Delta_{\mathbf{q}}$. Therefore, the condition that $F(\hat{\Delta}_{\mathbf{q}})<C(\hat{\Delta}_{\mathbf{q}})$ is sufficient for the validity of the mean field approximations. According to the Wick's theorem, the fluctuation of the superconducting gap is

\begin{equation}
\label{eq:fluc}
F(\hat{\Delta}_{\mathbf{q}})=\sum_{\mathbf{k},\mathbf{k'}}  \langle  L _{\mathbf{k}+\mathbf{q}/2,\mathbf{k'}+\mathbf{q}/2,\uparrow} L _{\mathbf{k}+\mathbf{q}/2,\mathbf{k'}+\mathbf{q}/2,\downarrow}   \rangle - \langle  M_{\mathbf{k}+\mathbf{q}/2,-\mathbf{k'}+\mathbf{q}/2, \uparrow} M_{-\mathbf{k}+\mathbf{q}/2,\mathbf{k'}+\mathbf{q}/2, \downarrow} \rangle.
\end{equation}

\begin{figure}
\centering
\includegraphics[width=0.8\textwidth]{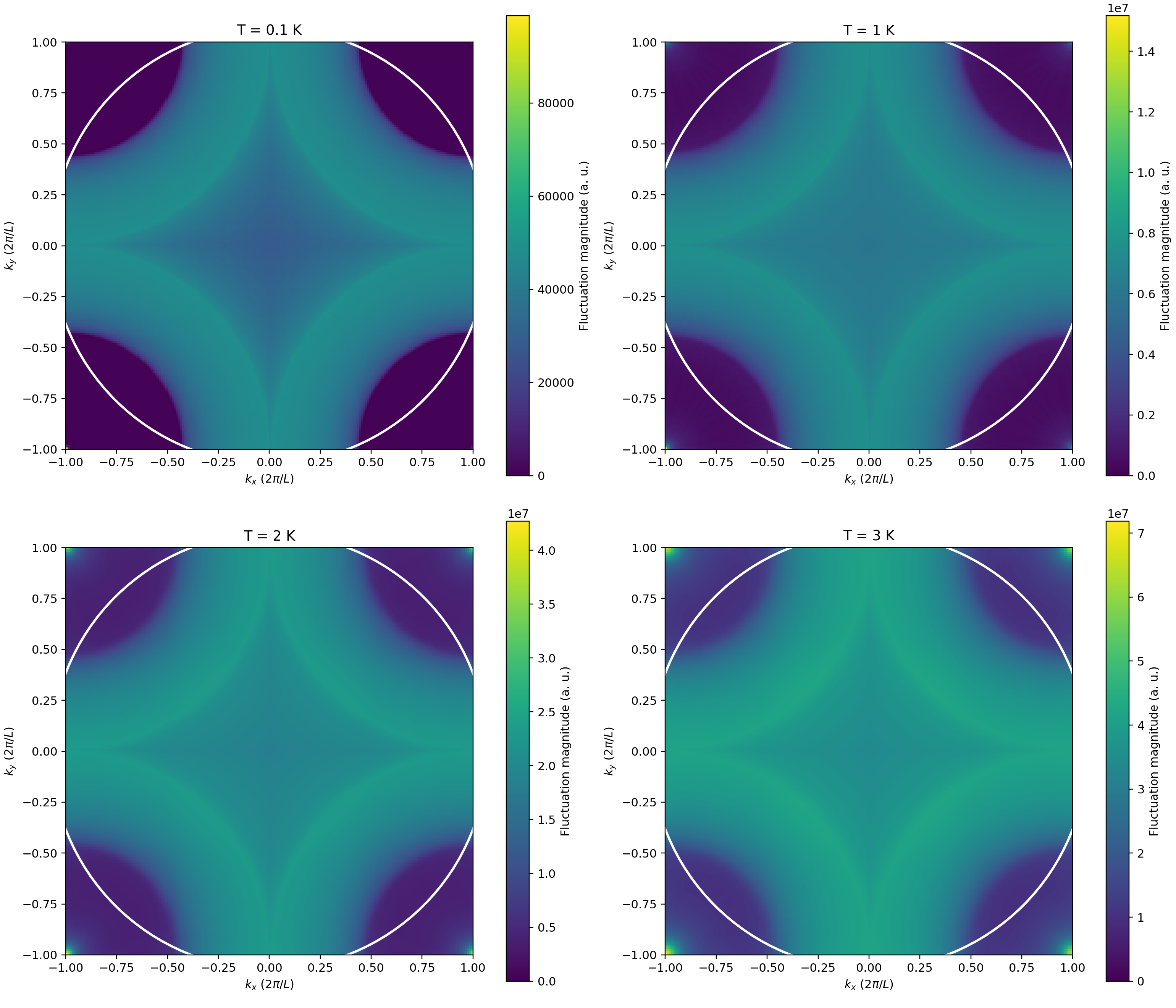}
\caption{The difference between the superconducting fluctuations and gaps is plotted with $k_x$ ($k_y$), the x(y)-component total wave vector of electron pairs at $   U=0.1 \mathrm{a. u.}, \mu = 0.01 \mathrm{a. u.}   $. The radius of the white circles is the critical wave vector at which the superconducting gap vanishes.}
\label{fig:fluc}
\end{figure}

\begin{figure}
\centering
\includegraphics[width=0.8\textwidth]{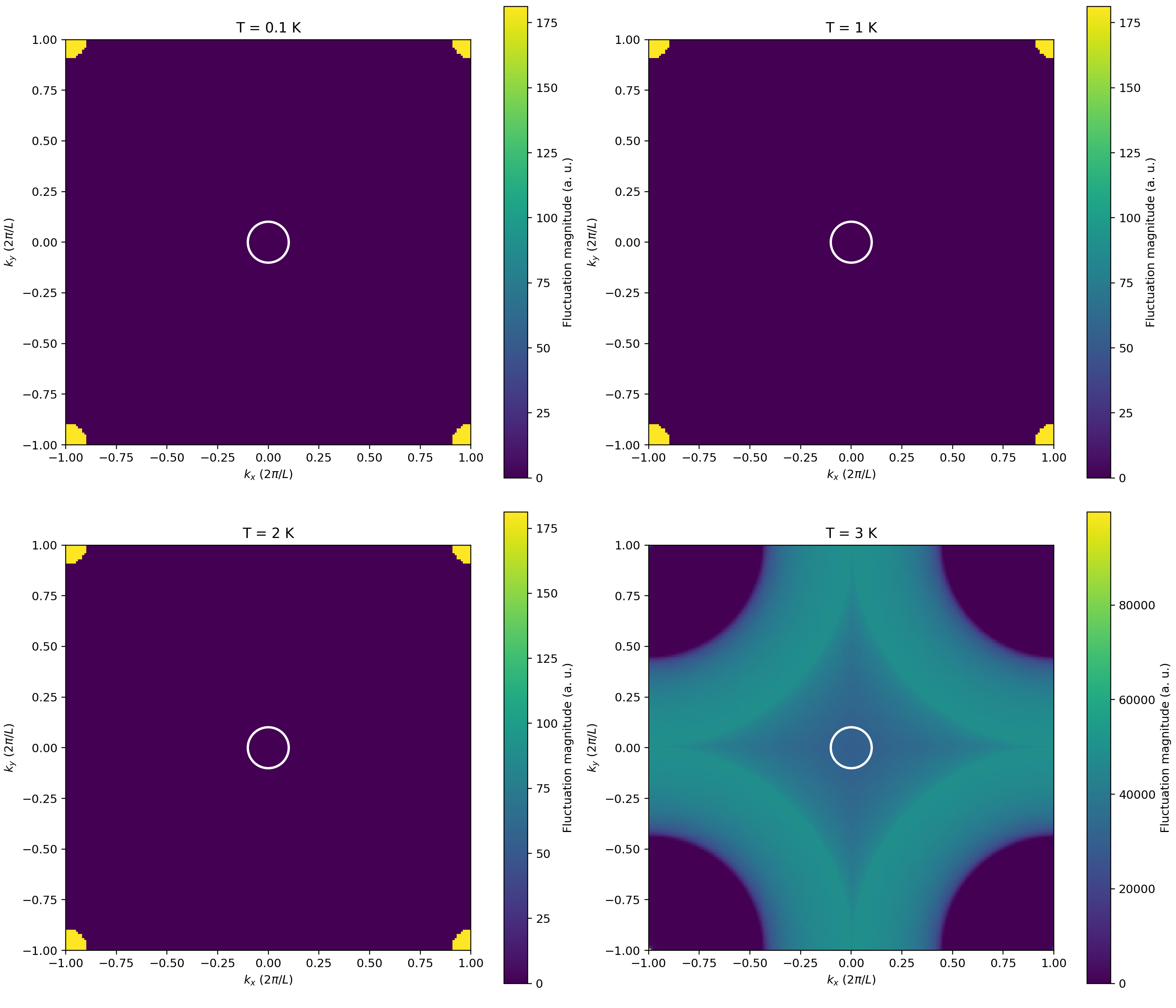}
\caption{The difference between the superconducting fluctuations and gaps is plotted with $k_x$ ($k_y$), the x(y)-component total wave vector of electron pairs at $  U=0.1 \mathrm{a. u.}, \mu = 0.05 \mathrm{a. u.}   $. The radius of the white circles is the critical wave vector at which the superconducting gap vanishes.}
\label{fig:fluc2}
\end{figure}

\begin{figure}
\centering
\includegraphics[width=0.8\textwidth]{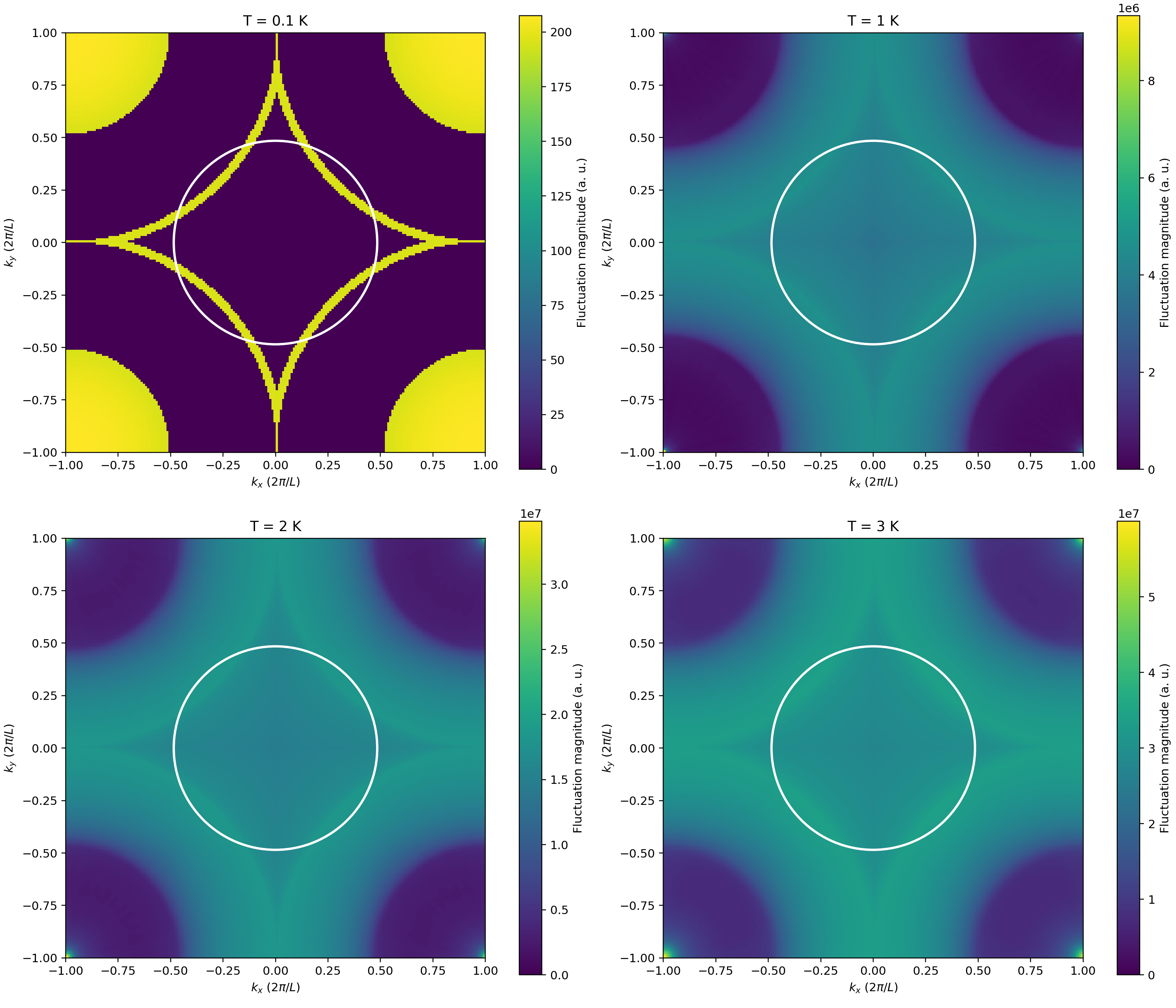}
\caption{The difference between the superconducting fluctuations and gaps is plotted with $k_x$ ($k_y$), the x(y)-component total wave vector of electron pairs at $  U=0.05 \mathrm{a. u.}, \mu = 0.01 \mathrm{a. u.}   $. The radius of the white circles is the critical wave vector at which the superconducting gap vanishes.}
\label{fig:fluc3}
\end{figure}

With the Eq. (\ref{eq:fluc}) and a square lattice geometry, the superconducting fluctuations minus the superconducting gaps at $U=0.1 \mathrm{a. u.}, \mu = 0.01 \mathrm{a. u.}$ are plotted for different temperatures. In Fig. (\ref{fig:fluc}), d-wave symmetry superconductivity appears at low temperatures, until fluctuations become large enough to destroy superconductivity and invalidate the validity condition of mean-field approximations at higher temperatures. By tuning the chemical potential from $\mu=0.01~ \mathrm{a. u.}$ to $\mu=0.05~ \mathrm{a. u.}$ (see Fig. (\ref{fig:fluc2})), the symmetry of superconductivity changes from d-wave to s-wave. Meanwhile, due to the presence of charge density wave order parameter, the superconducting fluctuations show several arcs in the Brillouin zone (see Fig. (\ref{fig:fluc3})). Limited by the computing power, we only searched the points with $U=1,0.5,0.1,0.05,0.01,0.005,0.001,0.0005~ \mathrm{a. u.},\mu=1,0.5,0.1,0.05,0.01,0.005,0.001,0.0005~\mathrm{a. u.}, T=0.1,1,2,3,5,7,9,11~\mathrm{K}$, and more sophisticated structures may still exist undetected. According to our calculations, for the parameter space $U=0.01 ~\mathrm{a. u.}, \mu = 0.01,0.005~ \mathrm{a. u.}  $, only s-wave superconductivity is observed, and for other different $\mu$, no superconductivity appears, which reveals the complicated properties of superconducting phenomena induced by the on-site Coulomb interaction in the intermediate coupling regime.

Generally speaking, the superconducting gap increases with the interaction potential $U$ and decreases with the temperature $T$. However, considering the influence of superconducting fluctuations, especially the charge density wave order parameters, the superconducting gap may not be observed when the mean field approximation is invalid. If the origin of superconductivity and the charge density wave order parameter is induced by the on-site e-e interactions, the superconducting and charge density wave order parameters will compete with each other, otherwise the two order parameters can coexist in the same sample. This argument can provide new insights into the origin of superconductivity in STO (electronic or phononic).

\section{Charge Density Wave Properties: Charge Density Wave Order Parameter, Fluctuations, and Effective Mass Enhancement}
\begin{figure}
\centering
\includegraphics[width=0.8\textwidth]{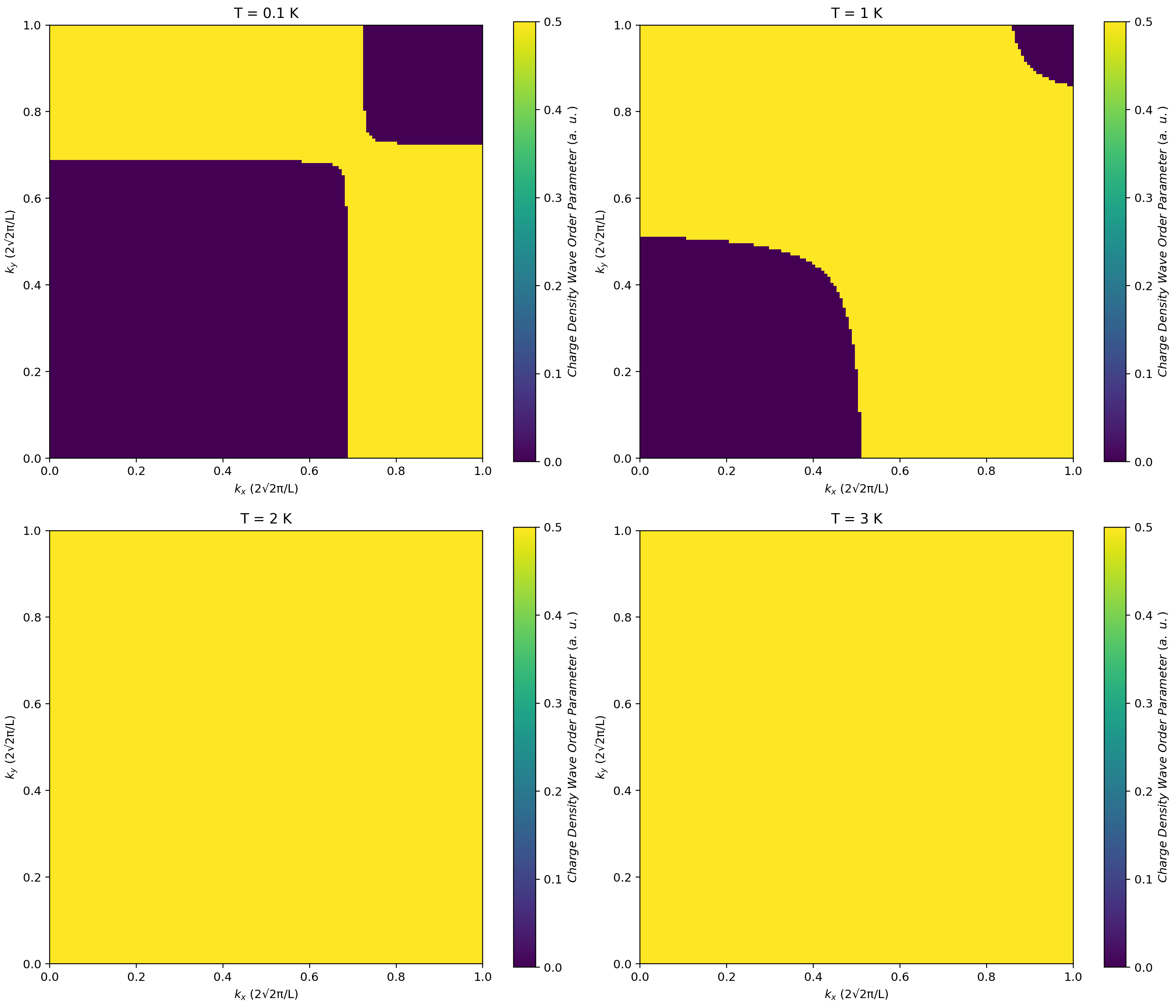}
\caption{The charge density wave order parameter is plotted $k_x$ ($k_y$), the wave vector magnitude of incident (scattered) electrons at $   U=0.01 \mathrm{a. u.}, \mu = 0.001 \mathrm{a. u.}   $.}
\label{fig:cdw}
\end{figure}

\begin{figure}
\centering
\includegraphics[width=0.8\textwidth]{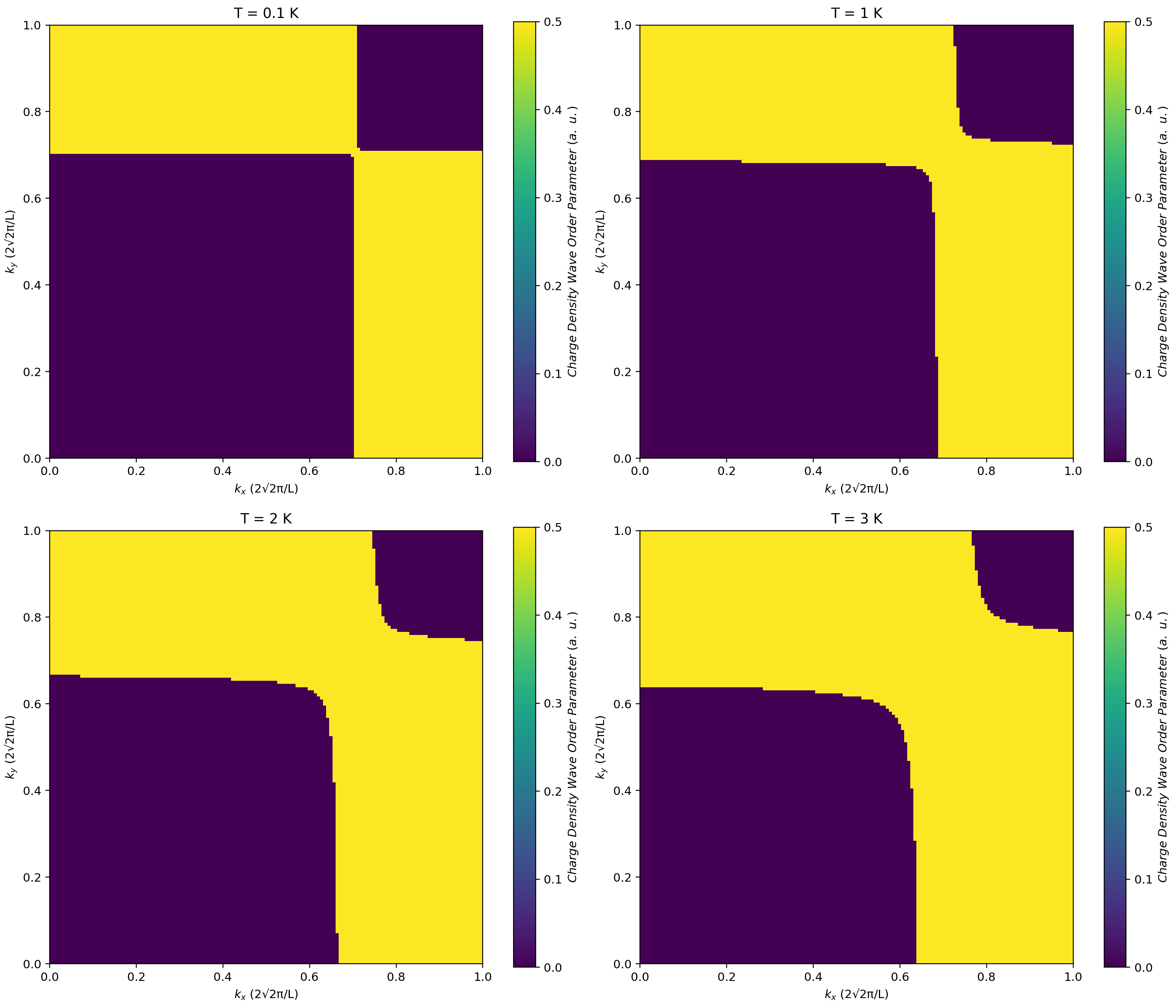}
\caption{The charge density wave order parameter is plotted $k_x$ ($k_y$), the wave vector magnitude of incident (scattered) electrons at $  U=0.1 \mathrm{a. u.}, \mu = 0.05 \mathrm{a. u.}   $.}
\label{fig:cdw2}
\end{figure}

Fig. (\ref{fig:cdw}) and Fig.(\ref{fig:cdw2}) depict the typical non-superconducting and superconducting states. In Fig. (\ref{fig:cdw}), the charge density wave order parameter induces large superconducting fluctuations and, therefore, destroys superconductivity. By contrast, the charge density wave order parameter has fewer non-zero elements and hardly affects the validity condition of the mean field approximations.

The fluctuations of the charge density wave order parameter, reduced by Wick's theorem is 

\begin{equation}
F(\hat{L}_{\mathbf{k},\mathbf{k'},\sigma}= N_{e,\mathbf{k},\sigma} N_{h,\mathbf{k'},\sigma}-|\langle \hat{c}_{\mathbf{k'},\sigma} \hat{c}_{\mathbf{k},\sigma}  \rangle|^2
\end{equation}

where $N_{e,\mathbf{k},\sigma} = \langle \hat{c}^\dagger_{\mathbf{k}\sigma} \hat{c}_{\mathbf{k}\sigma} \rangle, N_{h,\mathbf{k'},\sigma}= \langle \hat{c}_{\mathbf{k'}\sigma} \hat{c}^\dagger_{\mathbf{k'}\sigma} \rangle$ are the electron and hole particle number operators.$|\langle \hat{c}_{\mathbf{k'},\sigma} \hat{c}_{\mathbf{k},\sigma}  \rangle|$ is the thermal average of the triplet pairing superconducting order parameter, which assumes to be zero due to the absence of interactions between electrons with same spin directions. The validity condition of the mean field approximation for the charge density wave order parameter is $F(\hat{L}_{\mathbf{k},\mathbf{k'},\sigma}<C(\hat{L}_{\mathbf{k},\mathbf{k'},\sigma}$ where $C(\hat{L}_{\mathbf{k},\mathbf{k'},\sigma}=\langle |\hat{L}_{\mathbf{k},\mathbf{k'},\sigma}|^2 \rangle$. According to our calculations in the current parameter space, the total value of the charge density wave order parameter in the Brillouin zone increases with the rise of the interaction potential $U$ and temperature $T$, and the reduction of the chemical potential $\mu$. However, the decrease of the chemical potential $\mu$ can enlarge the density of states in a parabolic dispersion, and the increase of the temperature $T$ can facilitate the thermal excitations of electrons and holes. Both of the two circumstances will increase the value of $N_{e,\mathbf{k},\sigma} $ and $N_{h,\mathbf{k'},\sigma}$, and consequently invalidate the mean field approximations and establish an upper limit for the charge density wave order parameter.

The charge-density order parameter can increase the effective mass of electrons after diagonalization, combined with the kinetic part of the total Hamiltonian. If the increase in the effective mass is phononic, the electronic chemical potential $\mu$ can hardly affect the effective mass of electrons. Otherwise, the effective mass will increase as the electronic chemical potential decreases, provided that the increase is induced by on-site e-e interactions, as our calculations indicate. This argument offers a way to resolve the long-standing puzzles of the origins of electronic transport and the superconducting mechanism in STO (phononic or electronic). Previous research \cite{najev_electronic_2025} shows that the effective mass is independent of temperature and doping, which proves that the increased effective electronic mass is not induced by on-site e-e interactions.

\section{Magnetic Properties: Spin Density Wave Order Parameter, and Band Splitting Effects}

\begin{figure}
\centering
\includegraphics[width=0.8\textwidth]{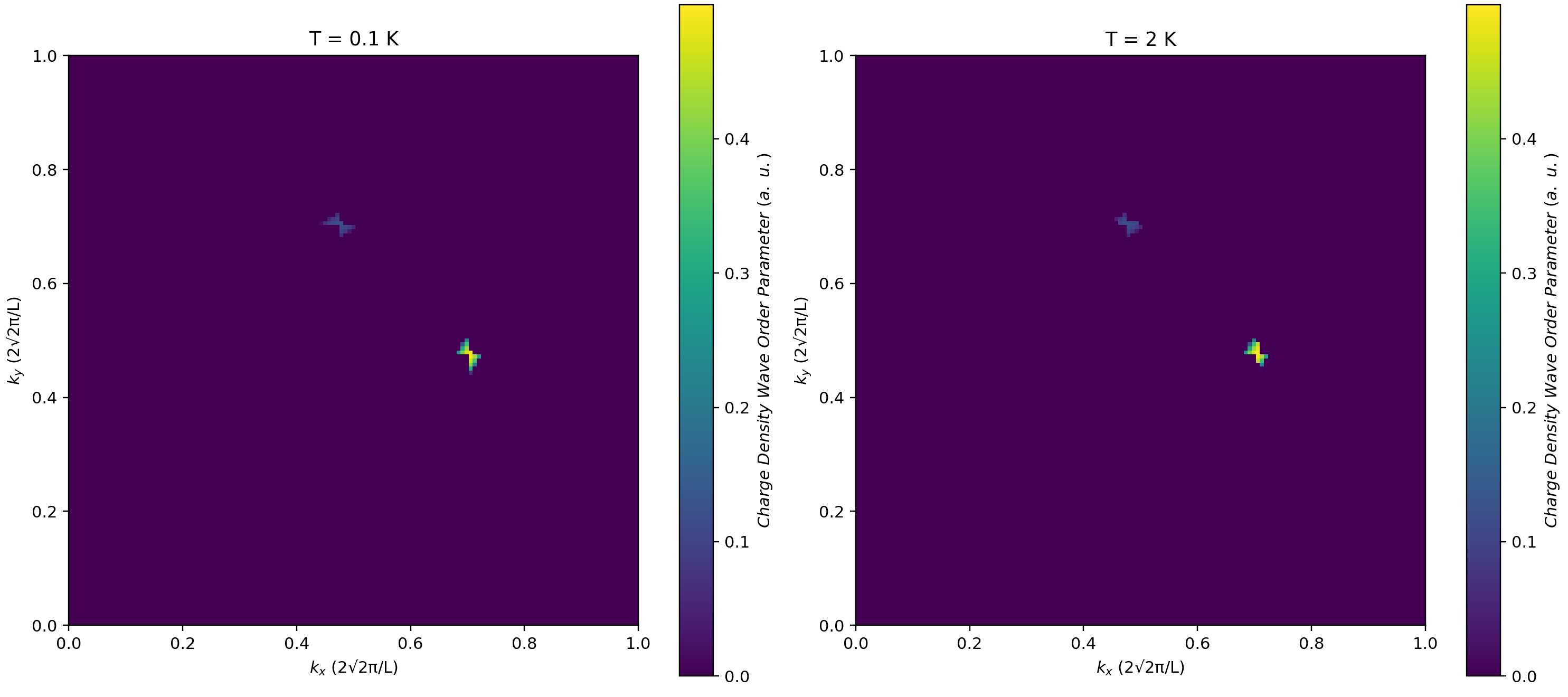}
\caption{The spin density wave order parameter is plotted $k_x$ ($k_y$), the wave vector magnitude of incident (scattered) electrons at $  U=0.01 \mathrm{a. u.}, \mu = 0.5 \mathrm{a. u.}   $.}
\label{fig:sdw}
\end{figure}

The spin density wave order parameters in the calculated parameter space are mostly zero, and the finite values appear randomly, even discontinuously, along with the change of temperature. In the rare cases when the spin density wave order parameter has finite elements (see Fig. (\ref{fig:sdw})), the other order parameters are identically zero, which indicates that the spin density wave order parameters compete with superconducting and charge density wave order parameters. Additionally, the spin density wave order parameter is sensitive to other order parameters, indicating that the spin density wave induced by on-site e-e interactions is difficult to detect.

Although the spin density wave order parameters are mostly zero, the commutation relations will yield an additional term that raises the energy of spin-up electrons (see Eq. (\ref{eq:sdw})). This term is not confined to the validity condition of mean-field approximations; it affects the system when other order parameters are negligible. Consequently, in most cases, the magnetic effects induced by the on-site e-e interactions are unnoticeable.

\section{Conclusions}
In this work, we have applied extended mean-field theory (eMFT) to the 2D Hubbard model describing degenerate dilute electron gases, focusing on elucidating the superconducting dome, fluctuations, and interaction mechanisms in strontium titanate (STO). By treating strong correlations non-perturbatively and validating the approach through fluctuation analysis, we demonstrate that on-site e-e interactions can induce a dome-shaped superconducting gap as a function of chemical potential, mirroring experimental observations in STO. Superconductivity exhibits d-wave symmetry at low doping, transitioning to s-wave symmetry at higher doping, with a critical wave vector that influences spectral features such as peak-dip-hump structures. Fluctuations, computed using Wick's theorem, confirm the regime of validity of mean-field approximations and reveal temperature-driven destruction of pairing.
Charge density wave (CDW) order emerges as a key competitor to superconductivity, inducing large fluctuations that suppress pairing in certain parameter regimes while enhancing effective electron mass, which scales inversely with chemical potential—a signature of electronic origins. Spin density wave (SDW) order is predominantly absent, appearing sporadically and discontinuously with temperature, underscoring its sensitivity and competition with other orders. An additional magnetic term from commutation relations subtly raises spin-up electron energies, potentially contributing to weak magnetization effects.
Our results highlight the intricate interplay of e-e correlations in dilute systems, providing new theoretical tools to resolve longstanding debates in STO: if effective mass enhancements or order-parameter competition are insensitive to the chemical potential, phononic mechanisms dominate; otherwise, e-e interactions play a significant role. These insights not only advance understanding of unconventional superconductivity in STO but also guide experimental probes, such as tuning doping near quantum critical points or measuring fluctuation spectra, to engineer enhanced $T_c$ in oxide interfaces and related materials. Future extensions could incorporate multi-orbital effects, disorder, or explicit e-ph coupling for a more comprehensive model.

\appendix
\section{A. Extended Mean-Field Theory with Multiple Order Parameters}
\label{app:sec1}
The algorithm of the extended mean-field theory is described as follows:
1. The Green's functions are defined as: $\Delta^\dagger_a=\langle \hat{c}^\dagger_{\mathbf{p},\uparrow}(\tau) \hat{c}^\dagger_{\mathbf{p'},\downarrow}\rangle$,$\Delta^\dagger_b=\langle \hat{c}^\dagger_{\mathbf{p},\downarrow}(\tau) \hat{c}^\dagger_{\mathbf{p'},\uparrow}\rangle$,$\Delta^\dagger_\uparrow=\langle \hat{c}^\dagger_{\mathbf{p},\uparrow}(\tau) \hat{c}^\dagger_{\mathbf{p'},\uparrow}\rangle$,$\Delta^\dagger_\downarrow=\langle \hat{c}^\dagger_{\mathbf{p},\downarrow}(\tau) \hat{c}^\dagger_{\mathbf{p'},\downarrow}\rangle$,$L_\uparrow=\langle \hat{c}^\dagger_{\mathbf{p},\uparrow}(\tau) \hat{c}_{\mathbf{p'},\uparrow}\rangle$,$L_\downarrow=\langle \hat{c}^\dagger_{\mathbf{p},\downarrow}(\tau) \hat{c}_{\mathbf{p'},\downarrow}\rangle$,$L_{h \uparrow}=\langle \hat{c}_{\mathbf{p},\uparrow}(\tau) \hat{c}^\dagger_{\mathbf{p'},\uparrow}\rangle$,$L_{h \downarrow}=\langle \hat{c}_{\mathbf{p},\downarrow}(\tau) \hat{c}^\dagger_{\mathbf{p'},\downarrow}\rangle$,$M_n=\langle \hat{c}^\dagger_{\mathbf{p},\uparrow}(\tau) \hat{c}_{\mathbf{p'},\downarrow}\rangle$,$M_r=\langle \hat{c}^\dagger_{\mathbf{p},\downarrow}(\tau) \hat{c}_{\mathbf{p'},\uparrow}\rangle$,$M_{h n}=\langle \hat{c}_{\mathbf{p},\uparrow}(\tau) \hat{c}^\dagger_{\mathbf{p'},\downarrow}\rangle$,$M_{h r}=\langle \hat{c}_{\mathbf{p},\downarrow}(\tau) \hat{c}^\dagger_{\mathbf{p'},\uparrow}\rangle$.

2. Calculate the imaginary time derivative of the above Green's functions using Heisenberg equations $\frac{\partial G}{\partial \tau}=\langle [\hat{H},\hat{d}(\tau)], \hat{d'} \rangle$ where $G=\langle  \hat{d}(\tau) \hat{d'} \rangle$ and transform the Heisenberg equations into frequency space.

3. Set random values for the elements in the frequency space of the 12 Green's functions defined in the first step.

4. Using the random values or the values from the last iterative round in the frequency space and the Heisenberg equations to alternatively calculate $L_{h\downarrow}$, $M_{hn}$, $L_{h\uparrow}$, $M_{hr}$, $\Delta^\dagger_a$, $\Delta^\dagger_b$, $\Delta^\dagger_\uparrow$, $\Delta^\dagger_\downarrow$,  $L_{\downarrow}$, $M_{n}$, $L_{\uparrow}$, $M_{r}$.

5. Transform the Green's functions $L_{\downarrow}$, $M_{n}$, $L_{\uparrow}$, $M_{r}$, $\$\Delta^\dagger_a$, $\Delta^\dagger_b$ into the imaginary time space, and compute the order parameters in the Hamiltonian $\Delta^{(\dagger)}_\mathbf{q} \equiv \langle\hat{c}^{(\dagger)}_{-\mathbf{k}+\mathbf{q}/2,\downarrow} \hat{c}^{(\dagger)}_{\mathbf{k}+\mathbf{q}/2,\uparrow}\rangle$, $L _{\mathbf{k}+\mathbf{q}/2,\mathbf{k'}+\mathbf{q}/2,\sigma} \equiv \langle \hat{c}^\dagger_{\mathbf{k}+\mathbf{q}/2,\sigma}  \hat{c}_{\mathbf{k'}+\mathbf{q}/2,\sigma} \rangle$, $M_{\mathbf{k}+\mathbf{q}/2,-\mathbf{k'}+\mathbf{q}/2, \uparrow} \equiv \langle \hat{c}^\dagger_{\mathbf{k}+\mathbf{q}/2,\uparrow} \hat{c}_{-\mathbf{k'}+\mathbf{q}/2,\downarrow} \rangle$, $M_{-\mathbf{k}+\mathbf{q}/2,\mathbf{k'}+\mathbf{q}/2, \downarrow}  \equiv  \langle \hat{c}^\dagger_{-\mathbf{k}+\mathbf{q}/2,\downarrow} \hat{c}_{\mathbf{k'}+\mathbf{q}/2,\uparrow} \rangle$ by taking the imaginary time to zero.

6. Examine whether the order parameters converge to constants. If they converge, stop the calculations and output the value of the order parameters. If the order parameters do not converge, go back to the 4th step and continue the iterative calculations.

We searched in the parameter space from $U=1 \mathrm{a. u.}, \mu=1\mathrm{a. u.}$ to $U=0.0001 \mathrm{a. u.}, \mu =0.0001\mathrm{a. u.}$ at different temperatures at $T =0.1, 1, 2 \mathrm{K}$ with the algorithm described above. However, if the mixing parameters are too small, the order parameters will diverge; if they are too large, the order parameters will continuously approach zero. We tried to set the initial values of the 12 Green's functions to larger values and increase the number of wave vectors to 200 $\times$ 200, but none of the methods yielded finite order parameters. The code for the algorithm is available upon request.

\section{B. Extended Mean-Field Theory with One Single Order Parameter}
\label{app:sec2}
The matrix form of Eq. (\ref{eq:del}) is 
\begin{equation}
H_\Delta = \sum_{\mathbf{k},\mathbf{q}} 
\begin{pmatrix} 
c_{\mathbf{k}+\mathbf{q}/2,\uparrow} & c^\dagger_{-\mathbf{k}+\mathbf{q}/2,\downarrow} 
\end{pmatrix} 
\begin{pmatrix} -E_{\mathbf{k}+\mathbf{q}/2} & -\frac{U}{2N}\Delta_{\mathbf{q}} \\ -\frac{U}{2N}\Delta^\dagger_{\mathbf{q}} & E_{-\mathbf{k}+\mathbf{q}/2} \\
 \end{pmatrix} 
 \begin{pmatrix} 
 c^\dagger_{\mathbf{k}+\mathbf{q}/2,\uparrow}  \\ c_{-\mathbf{k}+\mathbf{q}/2,\downarrow} .
 \end{pmatrix}
\end{equation}

The Bogoliubov transformation is implemented with the relations:

\begin{equation}
\begin{aligned}
\hat{c}_{\mathbf{k}+\mathbf{q}/2, \uparrow} &= u_{\mathbf{k'}} \hat{\gamma}_{\mathbf{k'} \uparrow} - v_{\mathbf{k'}} \hat{\gamma}_{-\mathbf{k'} \downarrow}^\dagger, \\
\hat{c}_{-\mathbf{k}+\mathbf{q}/2, \downarrow} &= u_{\mathbf{k'}} \hat{\gamma}_{-\mathbf{k'} \downarrow} + v_{\mathbf{k'}} \hat{\gamma}_{\mathbf{k'} \uparrow}^\dagger, \\
\hat{c}_{\mathbf{k} +\mathbf{q}/2,\uparrow}^\dagger &= u_{\mathbf{k'}}^* \hat{\gamma}_{\mathbf{k'} \uparrow}^\dagger - v_{\mathbf{k'}}^* \hat{\gamma}_{-\mathbf{k'} \downarrow}, \\
\hat{c}_{-\mathbf{k} +\mathbf{q}/2,\downarrow}^\dagger &= u_{\mathbf{k'}}^* \hat{\gamma}_{-\mathbf{k'} \downarrow}^\dagger + v_{\mathbf{k'}}^* \hat{\gamma}_{\mathbf{k'} \uparrow}.
\end{aligned}
\end{equation}

where $u_{\mathbf{k'}}=u^*_{\mathbf{k'}}=(\frac{1}{2}+x)^\frac{1}{2}$, $v_{\mathbf{k'}}=v^*_{\mathbf{k'}}=(\frac{1}{2}-x)^\frac{1}{2}$, $x \equiv \frac{E_{\mathbf{k}+\mathbf{q}/2}+E_{-\mathbf{k}+\mathbf{q}/2}}{2\sqrt{\frac{U^2}{N^2}\Delta_\mathbf{q}+(E_{\mathbf{k}+\mathbf{q}/2}+E_{-\mathbf{k}+\mathbf{q}/2})^2}}$. With the relations of the Bogoliubov transformation, the superconducting gap $\Delta_{\mathbf{k},\mathbf{q}} \equiv \langle \hat{c}_{-\mathbf{k}+\mathbf{q}/2, \downarrow} \hat{c}_{\mathbf{k}+\mathbf{q}/2, \uparrow} \rangle$ is rewritten as:

\begin{equation}
\Delta_{\mathbf{k},\mathbf{q}} = (\frac{1}{1+\exp(-\beta \lambda_1)}-\frac{1}{1+\exp(-\beta \lambda_2)})(\frac{1}{4}-x^2)^\frac{1}{2}
\label{app:d}
\end{equation}

where $\beta = 1/(k_B T)$, $k_B$, $T$ are the Boltzmann constant and temperature, and $\lambda_1=\frac{1}{2}(-E_{\mathbf{k}+\mathbf{q}/2}+E_{-\mathbf{k}+\mathbf{q}/2}-\sqrt{\frac{U^2}{N^2}\Delta_\mathbf{q}+(E_{\mathbf{k}+\mathbf{q}/2}+E_{-\mathbf{k}+\mathbf{q}/2})^2})$, $\lambda_2=\frac{1}{2}(-E_{\mathbf{k}+\mathbf{q}/2}+E_{-\mathbf{k}+\mathbf{q}/2}+\sqrt{\frac{U^2}{N^2}\Delta_\mathbf{q}+(E_{\mathbf{k}+\mathbf{q}/2}+E_{-\mathbf{k}+\mathbf{q}/2})^2})$

We set an initial value for $\Delta_\mathbf{q}$ and calculate the $\Delta_{\mathbf{k},\mathbf{q}}$ using Eq. (\ref{app.d}) and obtain $\Delta_{\mathbf{q}}$ with $\Delta{\mathbf{k}, \mathbf{q}}$ integrated over $\mathbf{k}$. The final $\Delta_{\mathbf{q}}$ will be output until the relative change is less than 0.01\%. The number of k-points is 200 $\times$ 200, which will not affect the output value of $\Delta_{\mathbf{q}}$. The code for the algorithm is available upon request.

\begin{acknowledgments}
This work is financially supported by the National Natural Science Foundation of China under Grant No. 12464009 and the Guangxi Natural Science Foundation under Grant No. AD21220127.
\end{acknowledgments}

\bibliography{bib.bib}

\end{document}